\documentstyle[twocolumn,aps,epsfig]{revtex}
\begin{document}
\title {
Possible Tau Appearance Experiment with Atmospheric Neutrinos
}
\author {
Todor~Stanev
}
\address {
Bartol Research Institute, University of Delaware,
             Newark, DE 19716, USA
}

\wideabs{
\maketitle
\begin{abstract}
\widetext
 We suggest an experimental measurement that could detect the
 appearance of tau neutrinos due to $\nu_\mu \rightarrow \nu_\tau$
 oscillations of atmospheric neutrinos by measuring the energy
 spectra of neutrino induced showers. $\tau$ neutrinos deposit a
 large fraction of their energy in showers generated by $\nu_\tau$
 CC interactions and the  subsequent $\tau$--lepton  decay.
 The appearance of $\nu_\tau$ will enhance the spectrum of neutrino
 induced showers in energy ranges corresponding to the neutrino
 oscillation parameters.  A shower rate lower than the `no oscillation'
 prediction is an indication for $\nu_\mu \rightarrow \nu_s$ oscillations.
\end{abstract}
}
\narrowtext

\section{Introduction}
  
  The Super--Kamiokande experiment, a densely instrumented
 water--Cherenkov detector of very large dimensions,
 confirmed~\cite{SuperK} earlier indications~\cite{IMBKam}
 that the abnormal ratio of atmospheric muon to electron neutrinos
 can be interpreted best as muon to tau or sterile neutrino
 oscillations. The oscillation hypothesis is supported by the low
 ratio of muon to electron neutrino events and by the disappearance
 of muon neutrino events as a function of the distance to their
 production site. Several independent data sets, measuring different
 neutrino energy ranges and neutrino interaction processes, are
 fully consistent with oscillations in maximum mixing and $\Delta m^2$
 value of 3.5$\times$10$^{-3}$ eV$^2$~\cite{SuperKnew}.
   
  None of the data sets, that have been currently analyzed,
 however, contain any signatures of tau lepton appearance.
 The low $\nu_\tau$ deep inelastic scattering cross section
 below 10 GeV and the very short $\tau$--lepton lifetime
 prevent the direct observation, although the experimental
 sample should contain a number of $\tau$--leptons. Several
 long baseline accelerator experiments have been proposed with
 the aim to detect $\tau$ neutrinos and confirm the $\nu_\mu
 \rightarrow \nu_\tau$ oscillation hypothesis.~\cite{LongB}. 
 These employ an accelerator neutrino beam that is more restricted
 in energy than the cosmic ray beam and high resolution near
 and distant ($\sim$700 km) detectors that are able to identify
 $\nu_\tau$ interactions.

  We propose an experiment that uses the short lifetime of the
 $\tau$--leptons to detect one of the  signatures of their
 appearance. Charge current $\nu_\tau$ interactions and the
 subsequent $\tau$ decays will create hadronic/electromagnetic
 showers that will practically coincide in vertex and in
 time. Much higher fraction of the $\nu_\tau$ energy will be
 deposited in the form of showers than in either $\nu_\mu$ CC
 interactions or in neutral current (NC) interactions
 of any neutrino flavor. A measurement of the energy spectrum
 of shower events initiated by atmospheric neutrinos will
 be able to register the appearance of $\nu_\tau$, that are 
 present at a very low level in the atmospheric neutrino flux
 in the absence of neutrino oscillations.~\cite{PasqReno}
 The shower signal is generated by atmospheric neutrinos
 of energy above 10 GeV. 
 The neutrino induced showers can be contained inside a big
 water (or ice) Cherenkov detector that does not have to be
 very densely instrumented. 

  In the case of $\nu_\mu \rightarrow \nu_\tau$ oscillations
 an excess of neutrino induced showers will created at certain
 shower energies that will reflect the values of the oscillation
 parameters. In the case of $\nu_\mu \rightarrow \nu_s$ neutrinos
 the shower rate will correspondingly decrease, although by 
 smaller amounts.
 
 \section{Showers generated by $\tau$ neutrino CC interactions}
 
  We envision a detector that consists of three or four strings of 
 photomultipliers in the fashion of the high energy neutrino
 telescopes~\cite{neutr_tel} on a circle of radius 10 to 15 meters
 instrumented with photomultipliers every 4 to 5 m.
 Such an arrangement has been already proposed~\cite{Moscoso} for the
 measurement of the energy spectrum of neutrino induced muons.
 This detector will not be able to reconstruct the shower development
 or differentiate between purely electromagnetic and hadronic showers,
 but should be able to contain showers of energy up to 1000 GeV.
 The depth of maximum X$_{max}$ for 1000 GeV electromagnetic
 showers is  $\sim$290 g/cm$^{-2}$ and about 100 g/cm$^{2}$ larger
 for hadronic showers, i.e. 3 -- 4 meters of water or ice. 
 Accounting for the absorption length of Cherenkov light, about
 25 m, the electromagnetic and hadronic showers will appear 
 indistinguishable in such relatively crude (in comparison to
 Super--Kamiokande) detector. So we define as shower energy $E_{sh}$
 the total energy released in the form of hadrons, photons and
 electrons in the final state, i.e. $E_{sh} = y \times E_\nu$
 in NC interactions. 
  
   Assuming that muon neutrinos and antineutrinos oscillate
 into $\nu_\tau$ ($\bar{\nu}_\tau$) there will be three sources
 of neutrino induced showers: those due to $\nu_e (\bar{\nu}_e)$
 CC interactions, NC interactions of all three neutrino flavors 
 and CC interactions of $\nu_\tau$ and $\bar{\nu}_\tau$. Electron
 neutrino CC interactions deposit the total neutrino energy in
 the form of a shower. The fraction of shower energy  in NC
 interactions is defined by the differential cross section
 $d \sigma/d y$ which is energy dependent. To determine
 $\nu_\tau (\bar{\nu}_\tau$) CC interaction contribution one
 has to add to $d \sigma/dy$ the fraction of the $\tau$--lepton
 energy $(1 - y) \times E_\nu$ which is not carried away by
 neutrinos at its decay. 

\begin{figure}[!hbt]
\centerline{\psfig{figure=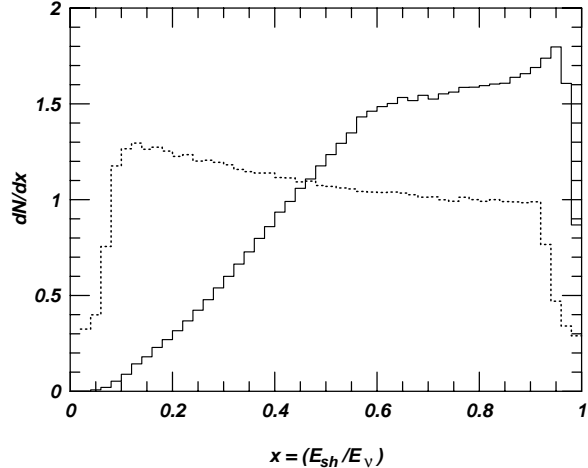,width=8.5cm}}
\medskip
\caption{ Fraction of the neutrino energy that converts to shower energy
 for $E_\nu$ = 10 GeV.
 The solid histogram is for $\nu_\tau$ CC interactions, dotted one
 is for NC interactions.
\label{fr-sh}}
\end{figure}

  Fig.~\ref{fr-sh} shows the distribution of the fraction of
 neutrino energy deposited in shower form for NC
 neutrino interactions and in CC $\nu_\tau$ interactions
 calculated using the GRV structure functions~\cite{GRV} and 
 $\tau$ decays performed by JETSET74~\cite{JETSET74}
 for E$_\nu$~=~10~GeV. The polarization of $\tau$ leptons is
 not taken into account. The two distributions are distinctly
 different. $E_{sh}/E_\nu$ peaks at 95\% in $\nu_\tau$ CC
 interactions with an average of 66\%. NC interactions yield  on
 the average 48\% of the neutrino energy in showers.

 \section{Neutrino fluxes for different oscillation parameters}

  In the energy range below 100 GeV atmospheric neutrinos are
 generated predominantly by the decay chain
 $\pi^{ch} \rightarrow \nu_\mu (\bar{\nu}_\mu) + \mu \rightarrow
 \bar{\nu}_\mu (\nu_\mu) + \nu_e (\bar{\nu}_e) + e^\pm $ of pions
 produced in interactions of cosmic rays in the atmosphere.
 In the 100 GeV range  the spectrum of neutrinos from pion decay
 is one power of $E$ steeper than the cosmic ray spectrum because
 of time dilation,  while those of neutrinos from muon decay are
 steeper by two powers of the energy.

 The oscillation probability in a simple two neutrino scenario
 is given in convenient units as
 $$ P_{\nu_1 \rightarrow \nu_2} = 
 \sin^2{ 2 \theta} \sin^2 \left[1.27 {{(L/ {\rm km}) ( \Delta m^2/ {\rm eV}^2)}
  \over {(E_\nu/ {\rm GeV})}} \right] \; ,$$
 where $\Delta m^2 = |m_{\nu_1}^2 - m_{\nu_2}^2|$ and $\theta$ is the mixing
 angle between the two mass eigenstates.

\begin{figure}[!hbt]
\centerline{\psfig{figure=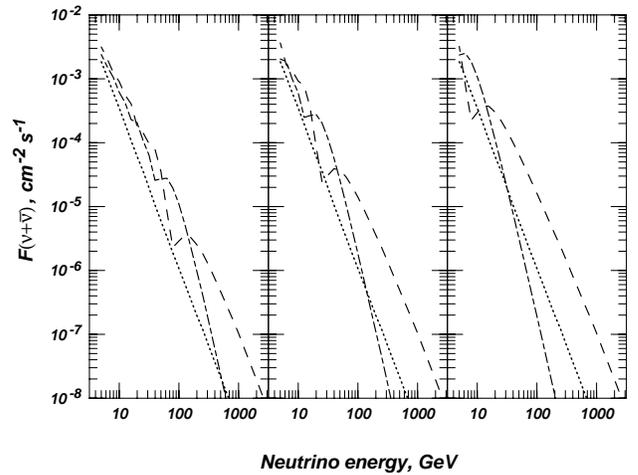,width=8.5cm}}
\medskip
\caption{Neutrino fluxes of all three flavors (dots: $\nu_e$, dashes: $\nu_\mu$,
         dash--dash: $\nu_\tau$) in the presence of $\nu_\mu \rightarrow
         \nu_\tau$ oscillations with maximum mixing and $\Delta m^2$ = 
          10$^{-2}$, 10$^{-2.5}$ and 10$^{-3}$ eV$^2$ from left to right.
\label{nu_fl}}
\end{figure}

  Fig.~\ref{nu_fl} shows the fluxes of atmospheric $\nu$+$\bar{\nu}$ for
 all three neutrino flavors coming from the lowest $\pi$ steradian of solid
 angle ($\cos {\theta}$ from -1 to -0.5) for maximum mixing and $\Delta m ^2$
 = 10$^{-2}$, 10$^{-2.5}$ and 10$^{-3}$ eV$^2$ derived from the atmospheric
 neutrino fluxes of Ref.~\cite{Agrawaletal}. The flux of muon neutrinos
 is now split between muon and tau neutrinos, the valleys in muon
 neutrinos corresponding to peaks in tau neutrinos. Because the
 neutrino pathlength in this $\cos {\theta}$ range varies only between
 R$_\oplus$ and 2R$_\oplus$ and the neutrino energy of interest 
 is relatively high, the oscillation patterns are not completely
 smeared, as they are in the GeV range. The peak in the $\nu_\tau$
 spectrum at 70 GeV and $\Delta m^2$ = 10$^{-2}$ eV$^2$, for example,
 corresponds to $\nu_\mu$ oscillation probability of 0.83 for 1 R$_\oplus$
 and 0.55 for 2 R$_\oplus$. The energy spectrum of $\nu_e$ is very
 steep and the $\nu_\tau$ flux (that derives from $\nu_\mu$) is higher
 than the $\nu_e$ flux below 500 (50) GeV for
 $\Delta m^2$ = 10$^{-2}$ (10$^{-3}$) eV$^2$. If the angular range were
 narrower and closer to vertical direction the peaks and valleys in
 the $\nu_\tau$ flux would be even more obvious.

 \section{The energy spectrum of neutrino induced showers}

\begin{figure}[!hbt]
\centerline{\psfig{figure=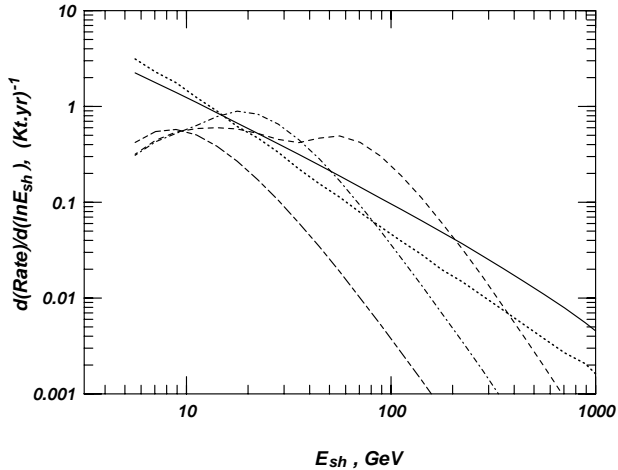,width=8.5cm}}
\medskip
\caption{ Shower energy spectra for: NC interactions (solid),
           $\nu_e$ CC events (dots) and $\nu_\tau$ CC events for 
           oscillations with maximum mixing and $\Delta m^2$ = 
           10$^{-2}$ (dashes), 10$^{-2.5}$ (dash--dot) and 10$^{-3}$
           (dash--dash) eV$^2$.
\label{sho_e}}
\end{figure}

 Fig.~\ref{sho_e} shows the energy spectra of showers generated
 by different neutrino flavors and interactions. The solid line
 indicates showers generated by NC interactions of all three
 flavors and the dotted line gives the showers of $\nu_e$ and
 $\bar{\nu}_e$ CC interactions. The sum of these two contributions
 is the expected shower energy spectrum in the absence of neutrino
 oscillations. The other three curves show the contribution of
 $\nu_\tau$ and $\bar{\nu}_\tau$ CC interactions for the three values
 of $\Delta m^2$ used in Fig.~\ref{nu_fl}. We assume here that
 the muons generated in $\nu_\mu$ CC interactions will be detected
 and used to veto the accompaning showers. 

 At the lowest shower energy shown, 5 GeV, the contribution of
 electron neutrino CC interaction is the biggest because $\nu_e$ deposit
 all of their energy in the form of showers. Because of the steepness
 of the $\nu_e$ energy spectrum, however, NC interactions dominate
 for shower energies above 15 GeV. The contribution of $\nu_\tau
 (\bar{\nu}_\tau)$ CC interactions is 5--10\% at 5 GeV. It increases
 with energy to become dominant for $\Delta m^2$ $\geq$ 10$^{-2.5}$
 and vanishes at high shower energy where the neutrino oscillation
 probability is very small. The low contribution at threshold is
 related to the small $\nu_\tau$ CC cross section at energies only
 slightly higher than the $\tau$ lepton mass.

\begin{figure}[!hbt]
\centerline{\psfig{figure=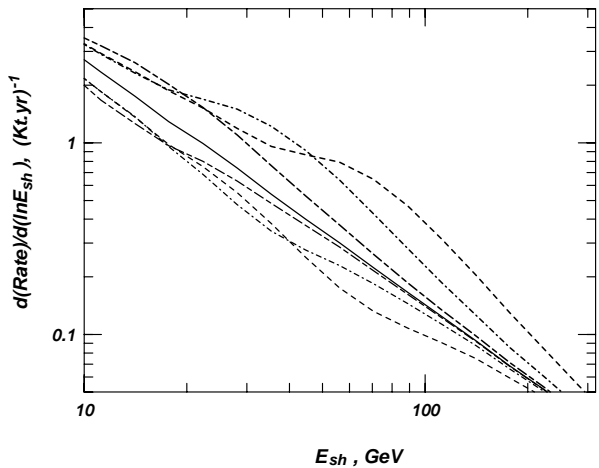,width=8.5cm}}
\medskip
\caption{ Shower energy spectra for
   $\nu_\mu \rightarrow \nu_\tau$ (thick lines) and
   $\nu_\mu \rightarrow \nu_s$ (thin lines) oscillations
   with maximum mixing  compared to the no oscillation case (solid line).
   $\Delta m^2$ = 0.01 eV$^2$ is shown with dashes, 0.005 eV$^2$ -- with
   dash--dot and 0.002 eV$^2$ -- with dash--dash).  
\label{sho_e_l}}
\end{figure}

 Fig.~\ref{sho_e_l} shows the total shower rate (the sum of all three
 contributions in Fig.~\ref{sho_e}) for $\Delta m^2$ of 0.002, 0.005
 and 0.01 eV$^2$. The no oscillation case is given with a solid line.
 The upper three histograms (thick lines) are for
 $\nu_\mu \rightarrow \nu_\tau$ oscillations while the lower three
 (thin lines) are for $\nu_\mu \rightarrow \nu_s$ case. The shapes of
 the energy spectrum that reflect the $\nu_\tau$ contribution are as
 essential as the  increase in the total rate. With the decrease of
 $\Delta m^2$ the shower excess moves to lower energy. The biggest
 excess for $\Delta m^2$ = 0.01 eV$^2$ is in the range of 40 to 150 GeV. 
 $\Delta m^2$ = 0.005 (0.002) eV$^2$ causes an excess above 20 (10) GeV. 

  In the case of $\nu_\tau \rightarrow \nu_s$ oscillations the changes
 in the shower rate are less spectacular. The only difference with
 the no oscillationscase is the missing contribution of NC interactions
 of the oscillated muon neutrinos. Still, the decrease of the rates 
 in the 10 -- 50 GeV range is of order 30 -- 40 \%.

 Table~\ref{tab1} gives the total rates of shower events in four
 wide logarithmically equally spaced energy intervals. The total
 rates above $E_{sh}$ = 10 GeV are 2.1 events per Kt.yr in $\pi$
 steradian in the absence of oscillations.
 This rate increases by 60\% to 3.35 for $\Delta m^2$ = 10$^{-2}$ eV$^2$
 in the case of $\nu_\mu \rightarrow \nu_\tau$ oscillations 
 and has intermediate values for lower $\Delta m^2$.
 The corresponding decrease in the case of sterile neutrinos 
 is by 25\% to 1.60.
 
\begin{table}[]
\caption{
    Shower event rates (in (Kt.yr)$^{-1}$) for different $\Delta m^2$
    in four energy bins for the lowest $\pi$ steradian of solid angle,
    i.e. between the nadir and 30 degrees below the horizon.
\label{tab1}}
\medskip
\begin{tabular} {r | c c c c }
$E_{sh}$, GeV & \multicolumn{4}{c}{ $\Delta m^2$, eV$^2$}\\ \cline{2-5}
 & 0 & 0.01 & 0.005 & 0.002 \\ 
 \tableline
 $\nu_\mu \rightarrow \nu_\tau$ & & &  \\
 10 -- 25  & 1.47 & 2.00 & 2.02 & 2.16 \\
 25 -- 63  & 0.46 & 0.88 & 0.98 & 0.64 \\
 63 -- 160 & 0.14 & 0.38 & 0.24 & 0.16 \\
 $ > $ 160 & 0.06 & 0.09 & 0.07 & 0.06 \\
 \tableline
 $\nu_\mu \rightarrow \nu_s$ & & &  \\
 10 -- 25  & 1.47 & 1.14 & 1.12 & 1.08 \\
 25 -- 63  & 0.46 & 0.31 & 0.30 & 0.41 \\
 63 -- 160 & 0.14 & 0.09 & 0.12 & 0.14 \\
 $ > $ 160 & 0.06 & 0.05 & 0.06 & 0.06 \\
\end{tabular}
\end{table}

 \section{Discussion and conclusions}

  The rates of shower events are sufficiently high for a relatively
 crude detector as described in the Introduction. Four strings 
 on radius of 15 m with length of 200 m each enclose a volume 
 of 90 Kt of water or ice. This translates into a statistics of
 190 events/yr above 10 GeV in the absence of oscillations and
 270 events/yr for $\Delta m^2$ = 0.002 eV$^2$.
 Two years of measurement should be enough to observe the
 enhancement in the shower rate with good statistics if
 $\Delta m^2$ is not lower than 2$\times$10$^{-3}$ eV$^2$.

  The requirements for the energy and angular resolution of
 the detector are not very high. Table~\ref{tab1} gives the
 rates in wide bins to demonstrate the relatively low
 sensitivity to the energy resolution of the detector.
 Smearing of the rates shown in Fig.~\ref{sho_e}
 with a Gaussian distribution mimicking  energy resolution of
 30\% does not change significantly the detectability of 
 $\nu_\tau$ appearance. Restricting the measurement to showers
 closer to vertical direction enhances the appearance effect
 as the peaks and valleys in the $\nu_\tau$ flux become more
 noticeable, although it decreases the statistics of the signal
 events. An angular resolution of order 10$^\circ$ should be
 good enough for the measurement of the rates in  $\pi$ steradian
 given in Table~\ref{tab1}.

  The presented calculation is not accurate enough to
 determine the actual detector rates because it assumes that:\\
 a) $\nu_\mu (\bar{\nu}_\mu)$ CC interactions are identified
 and are not counted as shower events. In practice some of
 these events (at high $y$) would not be distinguishable and
 will increase the experimentally measured shower rate;\\
 b) all $\tau$ decay energy not carried away by neutrinos
 contributes to $E_{sh}$. Actually a fraction of the $\tau$
 decay energy is carried by muons and does not contribute to
 the shower rate, especially in the $\tau^\pm \rightarrow
 \mu^\pm \nu_\mu (\bar{\mu}_\mu) \bar{\nu}_\tau (\nu_\tau) $
 channel with a branching ratio of 17.4\%~\cite{PDB}.\\
 These effects are not included in the calculation because their
 strength depends very much on the detector
 design and can only be studied by detector Montecarlo 
 codes that account for the Cherenkov light propagation in
 water or ice and detector efficiency.
  
  The absolute normalization of the predicted shower flux is not
 better than 20 -- 25 \%~\cite{Fratietal}. This does not, however,
 decrease significantly the sensitivity to oscillation parameters
 because the energy spectra of the shower events are significantly
 different for $\Delta m^2$ greater than about 2$\times$10$^{-3}$ eV$^2$.
 The spectral changes would be easier to detect if the measurement
 could be extended to shower energies lower than 10 GeV, where
 the contribution of oscillatated $\tau$ neutrinos is not significant.

 The large uncertainty in the absolute normalization will not
 allow a definite conclusion for the case of oscillations in sterile
 neutrinos. A combination of the muon neutrino disappearance with
 no increase of the shower rate will be, however, an important
 indication in favor of such oscillations.

 In conclusion, a measurement of the energy spectrum of neutrino
 induced showers will provide valuable complimentary information
 on the oscillations of atmospheric muon neutrinos. An increase
 of the shower rate would be a detection of the appearance of
 $\tau$ neutrinos in $\nu_\mu \rightarrow \nu_\tau$ oscillations
 for $\Delta m^2$ greater than 2$\times$10$^{-3}$ eV$^2$.\\[2truemm]
 {\bf Acknowledgements} The author acknowledges valuable discussions
 with R.~Engel, T.K.~Gaisser, F.~Halzen and D.~Seckel. This research
 is supported in part by U.S. Department of Energy contract DE~FG02~01ER~4062.

\end{document}